# Primary Beam and Dish Surface Characterization at the Allen Telescope Array by Radio Holography

G. R. Harp‡, R. F. Ackermann‡, Z. J. Nadler‡, Samantha K. Blair, M. M. Davis‡, M. C. H. Wright§,
J. R. Forster§, D. R. DeBoer§, W. J. Welch§

ATA GROUP

Shannon Atkinson‡, D. C. Backer§, P. R. Backus‡, William Barott‡, Amber Bauermeister§, Leo Blitz§, D.C.-J. Bock§, Geoffrey C. Bower§, Tucker Bradford‡, Calvin Cheng§, Steve Croft§, Matt Dexter§, John Dreher‡, Greg Engargiola§, Ed Fields§, Carl Heiles§, Tamara Helfer§, Jane Jordan‡, Susan Jorgensen§, Tom Kilsdonk‡, Colby Gutierrez-Kraybill§, Garrett Keating§, Casey Law§, John Lugten§, D. H. E. MacMahon§, Peter McMahon§, Oren Milgrome§, Andrew Siemion§, Ken Smolek‡, Douglas Thornton§, Tom Pierson‡, Karen Randall‡, John Ross‡, Seth Shostak‡, J. C. Tarter‡, Lynn Urry§, Dan Werthimer§, Peter K. G. Williams§, David Whysong

*Abstract*—The Allen Telescope Array (ATA) is a cm-wave interferometer in California, comprising 42 antenna elements with 6-m diameter dishes. We characterize the antenna optical accuracy using two-antenna interferometry and radio holography. The distortion of each telescope relative to the average is small, with RMS differences of 1% of beam peak value. Holography provides images of dish illumination pattern, allowing characterization of as-built mirror surfaces. The ATA dishes can experience mm-scale distortions across ~2 meter lengths due to mounting stresses or solar radiation. Experimental RMS errors are 0.7 mm at night and 3 mm under worst case solar illumination. For frequencies 4, 10, and 15 GHz, the nighttime values indicate sensitivity losses of 1%, 10% and 20%, respectively. The ATA's exceptional wide-bandwidth permits observations over a continuous range 0.5-11.2 GHz, and future retrofits may increase this range to 15 GHz. Beam patterns show a slowly varying focus frequency dependence. We probe the antenna optical gain and beam pattern stability as a function of focus and observation frequency, concluding that ATA can produce high fidelity images over a decade of simultaneous observation frequencies. In the day, the antenna sensitivity and pointing accuracy are affected. We find that at frequencies > 5 GHz, daytime observations > 5 GHz will suffer some sensitivity loss and it may be necessary to make antenna pointing corrections on a 1-2 hourly basis.

Manuscript received October 16, 2009. The authors would like to acknowledge the generous support of the Paul G. Allen Family Foundation, who have provided major support for design, construction, and operation of the ATA. Contributions from Nathan Myhrvold, Xilinx Corporation, Sun Microsystems, and other private donors have been instrumental in supporting the ATA. The ATA has been supported by contributions from the US Naval Observatory in addition to National Science Foundation grants AST-050690 and AST-0838268.

Authors indicated by an double dagger (‡) are affiliated with the SETI Institute, Mountain View, CA 95070 (phone: 650-960-4576, fax: 650-968-5830 e-mail: gharp@seti.org).

Authors indicated by a section break (§) are affiliated with the Hat Creek Radio Observatory and/or the Radio Astronomy Laboratory, both affiliated with the University of California Berkeley, Berkeley CA, USA.

*Index Terms*—Radio astronomy, dish surface accuracy, antenna radiation patterns, Allen Telescope Array, digital holography, image reconstruction, radio telescope, square kilometer array.

INTRODUCTION

The Allen Telescope Array (ATA) [1] is the first of a new class of LNSD (large number of antennas, small diameter) cm-wave interferometers and is outfitted with wide bandwidth receivers designed for the frequency range 0.5-11.2 GHz. Additionally the dishes use an offset Gregorian design (Figure 1) similar to ref. [2] which presents a nearly clear aperture for signal collection with low beam pattern sidelobes. At present the ATA comprises 42 offset Gregorian telescopes at Hat Creek Radio Observatory in Northern California, with plans to grow to 350 or more. Table 1 shows some of the antenna specifications and comparative as-built measurements associated with these specifications.

| Parameter | Spec. | Meas. X-pol | Meas. Y-pol |
|---|---|---|---|
| Pointing | 3' | 1.5' RMS Night 3' RMS Day | NA |
| Squint | No spec. | NA | 3.4' RMS |
| Dish Surface Accuracy | 3mm | 0.7mm Night 3mm Day | 0.7mm Night 3mm worst case Day |
| Sidereal tracking error budget, neglecting solar deformation | 12" = 10% synthetic beam width @ 10 GHz | 7.2" RMS (encoder step = 6") | 7.2" RMS (encoder step = 6") |

Table 1: Table of specifications for antenna pointing and dish surface accuracy compared with measured values.

The ATA is optimized for large scale SETI surveys and employing multiple points in the field of view (FOV) and for astronomical mosaic/on-the-fly imaging observations. The latter are wide FOV images pieced together from multiple pointings of the telescope. To stitch these fields together we require good characterization of the primary beam patterns from each of the 42 telescopes, and estimates of primary beam errors (e.g. solar heating deformations).

The ATA is an important precursor for the Square Kilometer Array (SKA) since it prototypes the LNSD architecture which is planned as a low-cost solution for SKA at frequencies greater than 1000 MHz [3]. The ATA uses a log-periodic feed which covers a decade in bandwidth [1]. Two competing telescope designs use 3-10 octave-band feeds on a carrousel to cover a wider frequency range, or multi-feed, phased-array and focal plane array feed designs [4], [5], [6], [7], [8]. Phased-array feeds cover a larger FOV than a single point feed, are typically optimized over about an octave frequency band. Though phased-array feeds capture a large area of the sky at one time, the average sensitivity of such feeds is a bit lower than a comparable single-point feed using similar components. Calibration of phased-array feeds is an area of intense study. Apart from receiver differences, two more major LNSD telescope arrays under development in Australia [9], [10] and in South Africa [11] will explore other realizations of the LNSD approach.

Because of ATA's unusual design, the antenna design community is interested in the accuracy and reproducibility of the ATA dishes, as well as their performance under solar radiation. Each dish is a single piece of hydro-formed aluminum (as opposed to separately adjusted panels) and such reflectors require novel mounting solutions. The wider community desires to learn what real-world performance can be achieved with this technology.

## A. Radio Holography

Microwave holography using distant radio emitters [12], [13], is an excellent tool for diagnosis of large diameter ($D \gg \lambda$) antennas and telescopes. Comprehensive measures of aperture fields, surface profiles, and feed alignment errors can be obtained from measured beam patterns and associated dish illumination images. Since its inception, substantial advances have been made [14], [15], [16], [17] and detailed applications of related radio holography methods abound (see e.g. [18], [19], [20], [21]).

Holography produces information about the coupled optical/ feed system, also comprising the feed focus (distance between the active area of the log-periodic feed and secondary focal plane). This complicated system is represented by the equivalent complex illumination pattern of the primary dish. Experimental results for the ATA show equivalent primary dish RMS surface errors with median 0.7 mm at night and 3 mm under worst case solar illumination (see section III for discussion of how this affects primary beam efficiency at different frequencies).

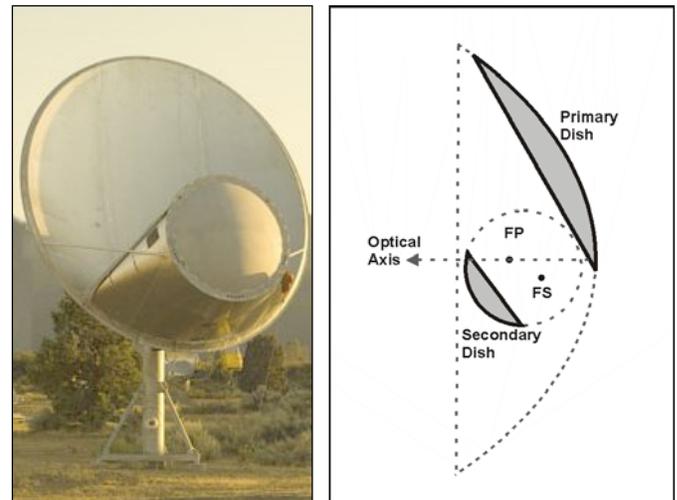

Figure 1: Left: Photo of an ATA dish during sunset. Right: Sketch of the ATA primary dish which comprises a section of a paraboloid. The face of the ATA dish is pitched at 29.4° relative to the face of a "fictitious" symmetrical dish (outer dashed curve) whose optical axis is shown. The primary dish has a focus at FP, which is refocused by the secondary to the point FS. Note the slight blockage of the primary by the top edge of the secondary. Using radio holography we characterize the primary dish, secondary dish, feed and supporting structures as a complex optical system. Artifacts arising from all optical elements are overlaid in this format, requiring close study to distinguish them. For instance, the ATA log-periodic feed requires some focusing depending on frequency, which must be considered.

## B. Beam Patterns

Besides holography results, this paper describes a thorough study of ATA beam patterns. The beam patterns, polarization, squint, and focus were measured with a 2-antenna interferometer technique using satellites and astronomical sources on most of the 42 telescopes.

We find that the ATA primary beam patterns are reproducible, with 1% RMS variation relative to beam peak from antenna to antenna at night (result of a quantitative analysis of the patterns in Figure 16. A squint of ~2% of the beam diameter (RMS) is observed at low frequencies (1.4 GHz). For a given antenna measured under the same conditions with 1-week time separation, the measurement reproducibility is 1 part in $10^5$ Equation (3) and ref. [22].

## II.   EXPERIMENT

We measure complex-valued beam patterns (telescope point distribution functions) by cross correlating the signals from antenna pairs. One antenna is pointed at a point source (satellite or CasA) and a second antenna is moved in a raster pattern about the source. The complex-valued correlation is averaged over a frequency band where the emitter transmits strongly and this represents the raw beam pattern measurement.

### A.   Target Choice

Figure 2 shows the frequency spectrum obtained from one antenna pair pointed at DSCS-III (defense satellite communication system), a geosynchronous radio satellite broadcasting near 7575 MHz. The vertical axis shows power in two linear polarizations and the shaded region indicates the frequencies used for collection of beam patterns. Direct observations of the polarization phase indicate that this satellite transmits dual-linear polarized radiation. Other satellites used for these measurements include Solidaridad F1 (RCP, geosynchronous, 1532 MHz), XM-2 (RCP, geosynchronous, 2335 MHz), Satmex-6 (dual linear, geosynchronous, 3877 MHz) and various Global Positioning System or GPS (RCP, maximal angular velocity roughly 3 times sidereal rate, 1575 MHz). The geosynchronous satellites were carefully chosen for frequency bands with stable emission and to be physically isolated from other satellites transmitting in the same band.

As indicated in Figure 2 only a sliver of frequencies (0.5 – 1 MHz) are integrated to produce the beam pattern on satellites. The emission is so strong that even this bandwidth produces signal to noise ratios (SNR) of $10^6$ in cross correlation.

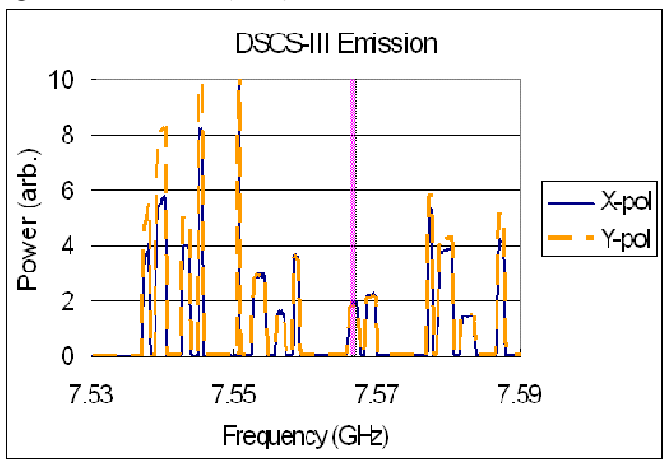

Figure 2: Power received from DSCS-III geosynchronous satellite showing most of its band for emission. The channels in the narrow magenta vertical shaded rectangle (7575 MHz) were selected for beam pattern measurement. This satellite can always be found in the vicinity of (Az, El) = (200°, 41°). The strongest features from this satellite appear at ~10 on this scale.

### B.   Beam Pattern Acquisition

While one antenna (1, reference antenna) tracks the target, the other antenna (2, moving antenna) executes a raster pattern (on the fly mapping[1]) about the target position. For each polarization, three correlations $V_{i,j}$ are used to compute the complex voltage beam pattern $A$ of antenna 2 in the instantaneous pointing direction as a function of azimuth and elevation:

$$A(\text{Az,El}) = \frac{V_{12}(t, \text{Az, El})}{V_{11}(t, \text{Az, El}) \; V_{12}^C(t_C, \text{Az}_C, \text{El}_C)}$$

where the indices refer to antenna number. The visibility for one point in the beam pattern $V_{12}$ is normalized to the autocorrelation of the fixed antenna $V_{11}$ to remove temporal variations in emission (note dependence on $t$). The phase of the central point is normalized to zero by dividing by the cross correlation when both antennas are pointed directly at the satellite $V_{12}^C$. $C$ stands for center This central point visibility is acquired rarely, only between every two raster lines.

Experimentally, there is an important difference between nearly stationary sources (like Geosynchronous satellites) and sources that move through large angles on the sky during data acquisition. The physical separation of the reference and moving antennas is ~100 m and this causes a time-varying time of arrival of the signal between antennas. For monochromatic radiation from a point source, the varying time of arrival introduces a nearly sinusoidal oscillation in the cross correlation between the two antennas. We call this sinusoidal variation "geometric fringe rotation." Geometric fringe rotation can be easily calculated and removed, and this is necessary for "moving" targets such as GPS or CasA.

With frequent returns to the center position, no fringe rotation correction is necessary to collect beam patterns on geosynchronous satellites. This was an important simplification in early ATA testing. On satellites we always measure narrow bandwidths (<1 MHz), but for beam patterns acquired on CasA we employ fringe rotation correction and integrate over a large frequency range (80 MHz) to improve sensitivity.

---

[1] The constant motion of the (horizontal raster) dish during observation introduces a small blurring of the beam pattern along the direction of motion. Since the pattern is collected at 3x Nyquist sampling, this blurring causes only a weak, unidirectional apodization or windowing effect in the aperture plane, which we ignore. If more scans were taken with a vertical raster, it would be possible to remove this effect.

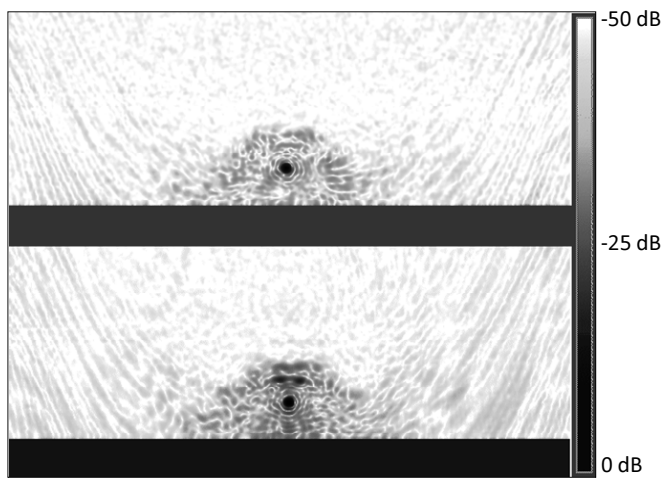

Figure 3: Power patterns at 2335 MHz showing half the sky visible to one antenna (X polarization top, Y polarization bottom). The gray scale varies over 60 dB in power units. In each image, the center of the darkest spot is the geosynchronous satellite XM-2, (Az, El) ~ (131.5°, 30°). The top of each image represents zenith and the bottom is minimum elevation for this antenna (17°). The horizontal scale represents 180 degrees in azimuth centered on the satellite azimuth. The detailed fine structure, up to 50 dB below the main peak, is reproduced in other measurements (not shown). The trapezoidal dark region surrounding the main beam is delineated by pointing angles where the feed has a direct line of sight to the target over the top of the secondary. The secondary occludes the feed line of sight at elevations greater than ~15° above the source.

The beam power was acquired for every azimuth (horizontal axis) and elevation (vertical axis) accessible by the antenna and displayed in a Mercator projection. The Mercator projection is one where azimuthal angles and elevation angles are the principle axes, and in some ways is easy to interpret, hence Mercator is often used to present world maps. Zenith of the observatory corresponds to horizontal line across the top, while the equator (elevation = $0^0$) is at the bottom of the black rectangle in each image. Despite the Mercator coordinates, zenith is not a degenerate position as a function of azimuth since the antenna presents a variable silhouette with change of azimuth. Hence we observe variations horizontally along the top of the image.

The black rectangular regions are not observed since the ATA dishes cannot point below 17° elevation. We show azimuthal angles (horizontal axis) between 90° and 270° relative to the satellite azimuth. Figure 3 displays half of a typical all-sky beam power pattern for one of the telescopes (full sky pattern requires about 2.5 days at 1 second per independent pointing). The gray scale spans 60 dB in power with contours at -20 dB, -40 dB and -60 dB.

The image was generated from the raw measurements of Eq. 1 by convolving each datum with a Gaussian with FWHM (full width at half maximum) equal to the HWHM (half width at half maximum) of the beam main lobe[2], and summing the results on a grid. While this sum is accumulating, a weighting grid is accumulated where each datum is replaced by unity. The Fourier transform of the data grid is divided by the Fourier transform of the weighting grid. This flat-fielding process eliminates variability due to non-uniform sampling (since raster lines are not necessarily synchronized).

To re-iterate, points below the target correspond to positions where the antenna points below the satellite. This helps us understand why there is a trapezoidal region of relatively high (-30 dB) sidelobes in the range ± 30° in azimuth and ± 15° in elevation surrounding the target. These are directions where feed spillover has a line of sight to the source as determined by inspection of the dish geometry (see Figure 1, right). Presumably if the dish could point lower (or the target were higher) the line of sight region would extend down to approximately 90° from the target, where the primary dish itself cuts off line of sight. It has also been suggested by one reviewer that our linearly polarized feed asymmetry could be partially responsible for this effect. On the optical axis (Figure 1, right), the secondary dish and ground shield block line of sight to the feed for elevations greater than about 15° above the target, in line with expectation.

In the Y polarization, two regions (the "dog's eyes") of higher sidelobe intensity appear (> -20 dB clearly evident in Figure 3 and Figure 4) approximately 10° above the target. Related but weaker features are observed in horizontally (X) polarized patterns (Figure 3). The origin of the dog's eyes is uncertain, though their compact size indicates that a large fraction of the primary reflector must be involved. We speculate that they might be caused by reflections from the primary dish entering directly into the backlobes of the exposed feed. To help realize this, notice that the primary dish focal point (FP) appears above and near the feed position (FS) in Figure 1. As the dish tips upward, the primary focus moves downward toward the feed position. Clearly, a full-scale modeling of the entire dish structure including the feed pattern, secondary, primary dish, and all support structures is indicated and necessary to understand such effects. The ATA team and collaborators hope to undertake such a study in the future (see discussion). The strength of the "dog's eye" features depends on frequency, focus, and polarization and is an ongoing area of research. We also note that these features are present even when the ground shield is removed.

Whatever their origin, these features are not associated with the dish surface errors since they would correspond to a strong, high-frequency ripple across one of the mirrors – evidently not present by inspection of the dishes and by the polarization dependence. Hence the measurements acquired to characterize the dish surface were restricted in solid angle to avoid these features. In future work the ATA team will undertake a systematic study of this feature, empirically or with quantitative simulations.

### C.   Coordinate Transformation

The data in Figure 3 are presented in a Mercator projection of azimuth and elevation. For holography, we re-grid the data onto a direction-cosine projection of the sphere where the target is at image center (see e.g. [12]). Though straightforward, this transformation has some interesting subtleties; for example, the holography formalism cannot handle waves arriving from behind the dish. Details are described in [22]. Figure 4 shows the data after re-gridding where we have selected data within 30° of the target (great circle angle).

---

[2] Convolution of the beam has the impact of down weighting the edges of the holographically reconstructed dish image. This down weighting is corrected in the dish images below simply by dividing the final dish image by the Fourier transform of the convolution function.

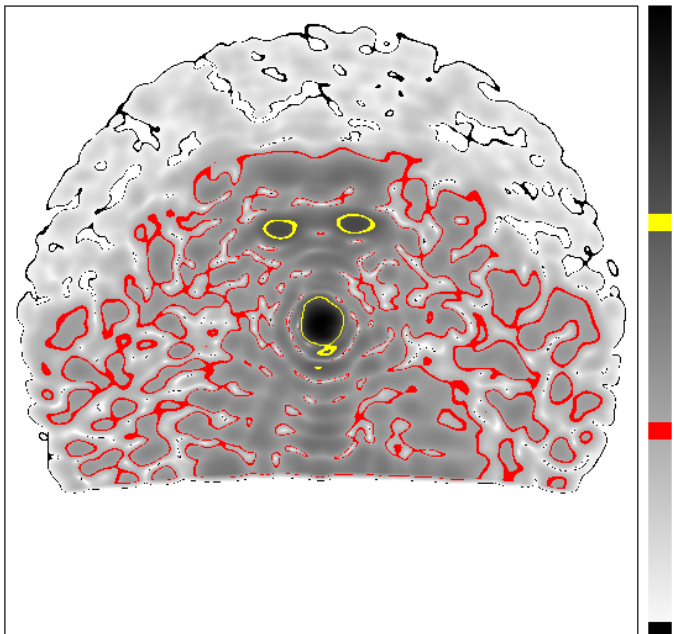

Figure 4: The Y-polarization data of Figure 2 are re-gridded into a coordinate system linear in the relative angle between the beam pattern center and each measured point. Beam power over 60 dB as in Figure 2 with contours at 0, -20, -40, and -60 dB. Notice that the lower horizon (bottom of image, elevation = 17°) is rendered as a curved line due to this non-linear transformation.

Figure 4 displays the beam power. There are ring-like sidelobes surrounding the main lobe of the beam pattern; these are related to the Airy rings of a circular aperture. The data are convolved with a cosine function that decreases smoothly from 1 to zero over an angular range of half the width of the main lobe (here, about 1°). Apodization [23] in the measurement domain before performing a Fourier transform is a well known method to reduce unphysical "edge" artifacts caused by the sudden drop to zero at the edge of the measured beam pattern.

The logarithmic scale used to display these images greatly enhances the appearance of far-out sidelobes which contribute very little to the received signal. (For figures showing the shape of the main beam pattern on a different scale, see Figure 16). Up to the second sidelobe, the sidelobe patterns are similar from dish to dish (with some exceptions), and vary slowly with frequency. In Figure 4, the "far out" sidelobes are not reproducible, and are presumably dominated by slight differences in dish mounting compounded by multi-path scattering between the ground shield ("shroud") and support hardware, and from deformations in mirrors which vary from antenna to antenna.

## D. Holographic Reconstruction

The complex-valued beam voltage pattern can be Fourier transformed to reveal the primary dish illumination pattern [12, 13]. Although we refer to the primary dish only, we emphasize that such illumination patterns include displacements and deformations of the secondary dish as well as displacements, mispointing or squint of the receiver (see Table 1). Because the radio wavelength to dish diameter ratio (λ/D) is large by comparison to most radio telescopes, it is important to re-examine the theoretical basis for this kind of holography to be sure that no implicit assumptions are violated. The wide-angle corrections we describe here are very similar to corrections discussed in [24].

For example, the Helmholtz-Kirchoff integral (which is essentially a truncated three dimensional Fourier Transform) gives the electric field amplitude $A(\vec{r})$ at a position $\vec{r}$ from the field amplitude $A(\vec{k})$ in the direction $\hat{k}$ and at a frequency corresponding to $k = \dfrac{2\pi}{\lambda}$ using

$$A(\vec{r}) = \int_{\hat{k}_{\text{Measured}}} A(\vec{k}) \exp(-i\vec{k} \cdot \vec{r})\, d\hat{k} . \qquad (2)$$

Since the radiation is nearly monochromatic and the far field sky is a spherical shell, we have used $\vec{k} = k\hat{k}$. The angular range of beam pattern measurement is essentially determined by the desired spatial resolution over the reflector/aperture.

On a large dish (small λ/D) it is customary to measure the beam pattern over only a small angular range (<<10°) on the sky close to the primary beam maximum. Since the $z$-component of $\vec{r}$ may be written $z = k\big(1 - \cos(\theta)\big)$, when $\theta \ll 1$, $z \cong 1$ and this term can be neglected to give a two dimensional integral.

On dishes like those at the ATA, λ/D is larger than for e.g. VLA (Very Large Array, Socorro, NM) dishes, which means that beam patterns must be acquired over relatively large angles (10°) to obtain several fringes (several primary beam widths) for holographic transform. For this reason, we re-map the raw beam pattern values from elevation and cross elevation angles to a direction cosines grid (projection of sphere onto a plane) to prevent distortions at the edge of the beam pattern.

While evaluating our implementation of this integral, regridded data like that of Fig. 4 were used. Beam pattern points less than 90° from the pointing center are included in a discrete Fourier transform according to equation (2) to give the raw dish image. At the same time, a second image of the "dish weight" is accumulated where $A(\vec{k})$ is replaced with unity. The dish images are apodized with (multiplied by) a Gaussian function with FWHM = 3 m (corresponding to a Gaussian convolution of the beam pattern). The raw dish and weighting images are Fourier transformed to the beam plane, where the raw beam pattern is normalized to the beam pattern weight. This step gives uniform weight to each pattern point. The beam maximum might be off-center if the antenna pointing-model is imperfect[3]. The mispointing (as in e.g. Figure 10) is estimated and the dish image is recomputed with an inverse FFT. The mispointing is then removed in the dish plane using a 2D phase gradient and the 3 m FWHM Gaussian apodization is removed by division in the dish plane, resulting in the final dish image. A forward FFT shows the final beam pattern. All the images shown here were analyzed in this way.

There is one final analysis used in computing the root mean square (RMS) error of the dish surface. The amplitude weighted phase pattern of the dish is fit with a quadratic polynomial in phase which simulates a geometrical mis-focus term $\exp(i\kappa r^2)$ in the dish plane. This focus term is removed from the dish phase pattern before calculation of the RMS error. This was important to success of the project because we don't know the optimal focus *a priori*, though it is determined empirically later in the study.

### E.    Beam Patterns

Nighttime measured beam patterns from ATA antennas are stable over periods of weeks at a time down to one part in $10^5$ of the primary beam maximum [22].

$$A(\vec{k}, t_0 + 1)\,\text{week} = (1 \pm 10^{-5})\, A(\vec{k}, t_0),\qquad (3)$$

and are reasonably fit with a 2-D circular Gaussian beam. This point is supported by an analysis where a 2-D Gaussian is fit to 99 beam patterns at different frequencies down to the 10% and 50% power levels. Because the beam FWHM should be inversely proportional to frequency, it is convenient to define the relationship

$$\text{FWHM}(f_{GHz}) = \text{FWHM}(1\ \text{GHz}) / f_{GHz}\qquad (4)$$

which gives a rule of thumb for the FWHM at any frequency. For the 1% power level measurements, the best fit for the frequency dependent beam FWHM as $3.7°/f_{GHz}$. Fitting the beam patterns to the 50% power level gives a rule of thumb of $3.5°/f_{GHz}$. Ordinarily we use the latter equation and the 5% difference is an indication of the variance of the actual beam shape from a true Gaussian beam.

Knowing this fact, we expect that very high dynamic range imaging in mosaics at the ATA will require, at minimum, a more sophisticated analytical function for the beam shape. The true beam shape is important because mosaics are stitched together such that their edges overlap at say, the 25% power level. Sources at the edge of the beam are strongly affected by small primary beam and model differences where a 10% difference corresponds to a 40% error in the observed flux. To generate optimally accurate mosaics, a better analytical function has to be developed or we might use empirical primary beam patterns for every antenna. At present, neither of these features are supported in some radio astronomy analysis packages such as MIRIAD [25].

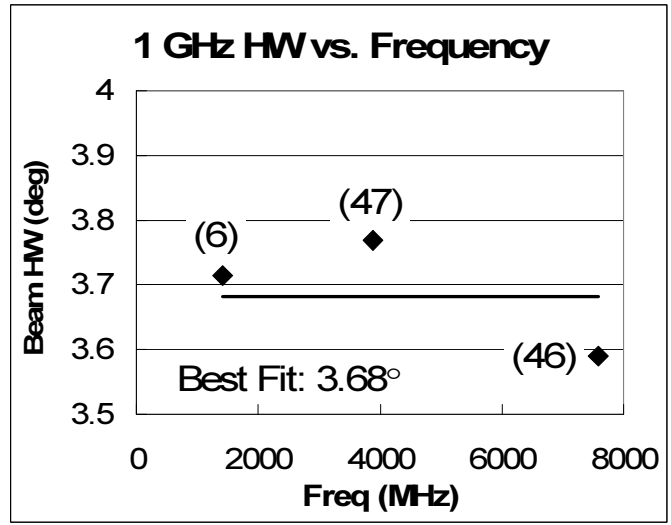

Figure 5: The measured best fit Gaussian half width averaged for many antennas and polarizations (approx. equal X and Y polarizations) at three frequencies. These results show the fit down to the 10% beam power level. Parenthetical numbers indicate how many antennas contributed to each point in the plot. Each half-width value has been scaled by the ratio of the observation frequency to 1000 MHz, which ideally puts all values on a constant line (black line) with a single parameter: Half Width = $3.7° / f_{GHz}$. The same fitting using beam power patterns down to the 50% power level give the average result Half Width = $3.5° / f_{GHz}$.

To further demonstrate the character of the ATA primary beam, we have computed average beam power patterns using both X and Y polarization data (same beam patterns included in Figure 5). These averaged patterns are shown on a linear scale in both grayscale and contours in Figure 6. The 1420 MHz pattern exhibits a weak noise background since it was obtained on a relatively weak source (CasA) as compared to 3877 and 7575 MHz which were obtained on satellites. No attempt to remove the background noise level was performed on any of the beam pattern images.

---

[3] Because of the squint, it is impossible to have a pointing model that is accurate for both polarizations. Typically we optimize for the x- or horizontal-polarization. Since the y-pointing is thought to be offset from optimal (feed droop), this same effect may be responsible for the "dog's eyes." Future simulations may bear this out.

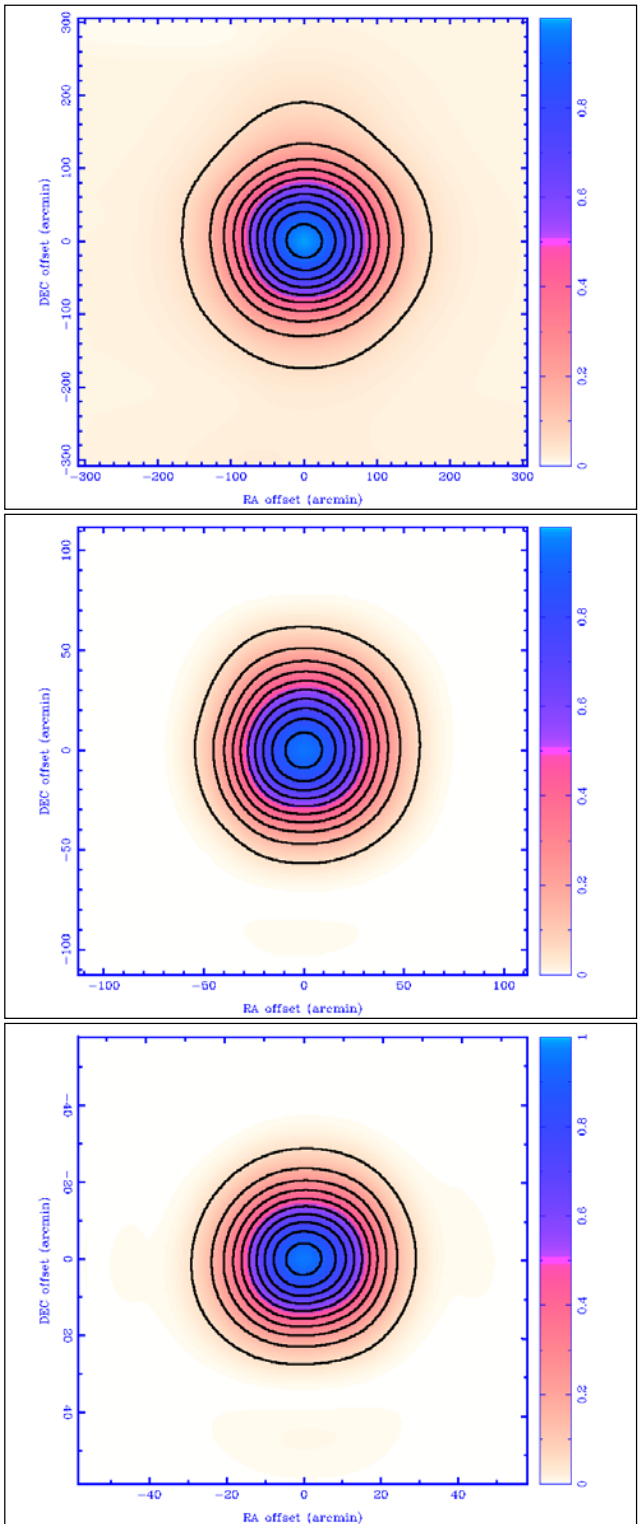

Figure 6: Averaged beam power patterns on a linear scale. Frequencies are, from top to bottom, 1420 MHz (CasA), 3877 MHz (Satmex-6), and 7575 MHz (DSCS-III). Contours at 90%, 80%, 70%, 60%, 50%, 40%, 30%, 20%, 10%. The angular width of the images have been scaled proportional to observing wavelength so that contour positions can be compared directly.

Figure 7 shows more detail with line cuts through a single beam pattern taken at 3877 MHz in the horizontal and vertical directions. One can see that the dish has mirror symmetry about a vertical mirror plane, and as expected the horizontal cut through the beam pattern shows good symmetry about boresight. Likewise, the dish is asymmetric across any horizontal mirror plane, leading to an asymmetric beam shape on the vertical cut. The vertical cut shows a slight shoulder above boresight near the half power point, and a matching dip below. This shoulder and dip is present in beam patterns at all frequencies, and there are variations in the magnitude of the shoulder (dip) depending on frequency, as seen in Figure 8.

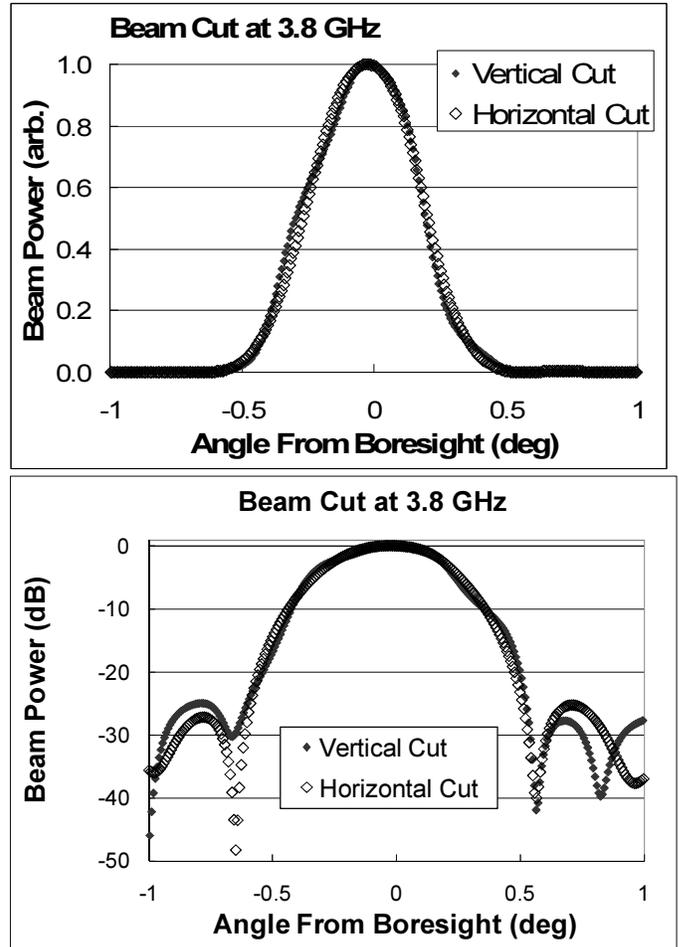

Figure 7: Two plots showing the same beam power data through a single antenna beam pattern. The open diamonds are the horizontal cut and black diamonds are the vertical cut. The horizontal beam cut is fairly symmetrical about zero angle, consistent with the symmetry of the antenna. Likewise the vertical cut is asymmetric due in part to a slight blockage of the primary antenna by the offset secondary dish.

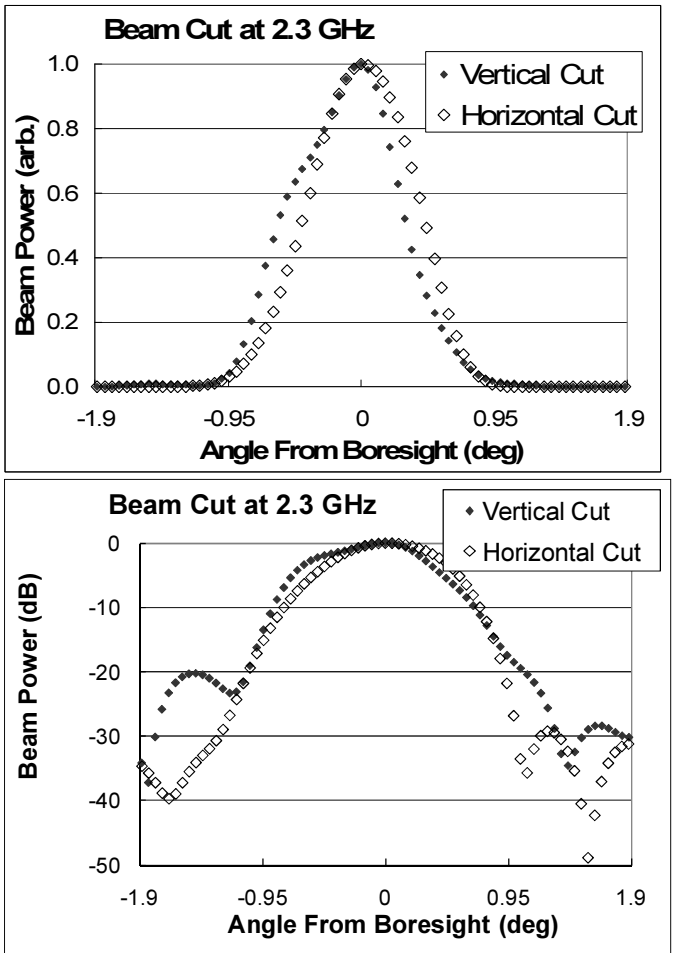

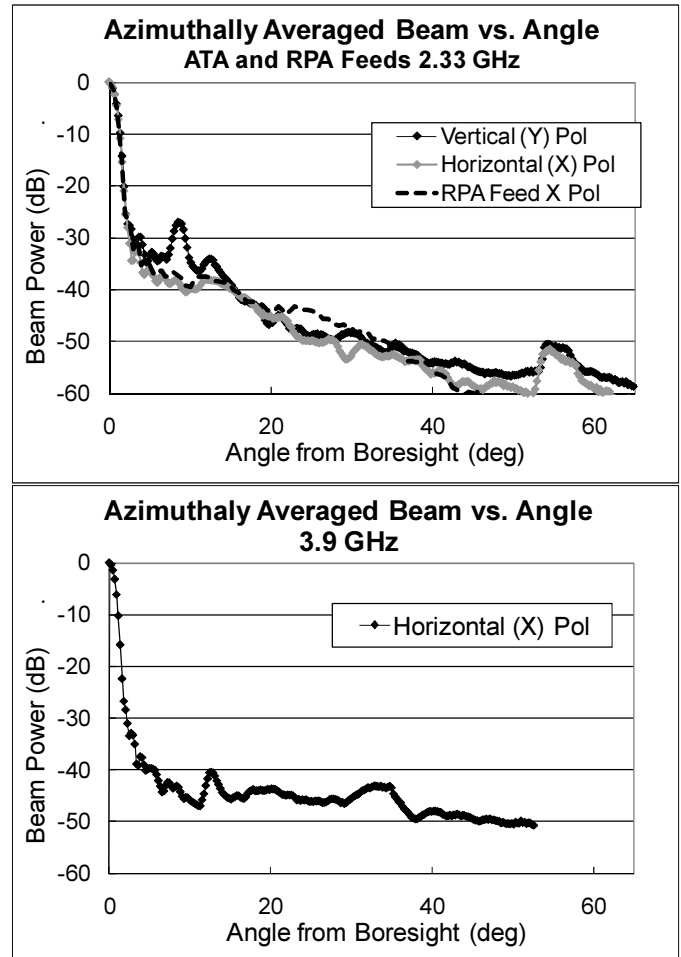

Figure 8: Similar to Figure 7 but for a different antenna and at a different frequency (2334 MHz). Qualitatively, this pattern shows the same symmetry properties of the antenna in Figure 7, with a fairly symmetrical pattern on the horizontal axis and a shoulder and dip in the vertical axis.

Figure 9: Azimuthally averaged beam power as a function of boresight angle at 2335 (top) and 3877 MHz (bottom). At top, vertical polarization (diamonds) data are taken from the same data as in Figure 8 and the horizontal polarization data (gray squares) were acquired at the same time. The dashed line is data taken with a corrugated horn feed in the X polarization (dashed line, RPA feed). The bottom curve shows data taken at 3877 MHz on X polarization using an ATA feed.

These graphs indicate another way that the ATA primary beam pattern is not a perfect match to a 2-D Gaussian beam profile. Again, this deviation from Gaussian behavior suggests that optimal dynamic range in mosaicked images will be achieved only with more sophisticated modeling of the primary beam. One may ask, how perfect does the fit have to be? The answer depends on the desired image fidelity. The image fidelity is limited by the ratio of the strongest deviation of the true beam pattern from the analytical or numerical model. As a rule of thumb, a figure of merit (FOM) maybe defined: FOM = RMS(abs($1 - A_{model} / A_{real}$)), which is about 1% for the simple Gaussian models implemented now in the standard ATA reduction software. With this FOM, we expect about 1% image fidelity for any given spatial frequency in the image and optimistically as low as $1\% / \sqrt{N_{Ant}}$ for the image produced from all baselines. As a standard, notice that the square kilometer array (SKA) is shooting for $10^{-4}$ image fidelity.

Another way of looking at the beam pattern is to get a statistical perception of the beam sidelobe level as a function of boresight angle out to very large angles. This is shown in Figure 9 where we plot the azimuthally-averaged beam power (average power of a unit flux point source) for a number of situations. In all cases, the average sidelobe level decreases almost monotonically with increasing boresight angle, falling below -40 dB for angles greater than 17° at 2335 MHz and angles greater than 5° at 3877 MHz.

The peaks in Figure 9 in the range 10-15° are related to the "dog's eyes" in Figure 4. The peaks at 50-60° at top and 30-40° at bottom are artifacts caused by measurements at the telescope horizon where strong RFI can cause >-50 dB "false" correlations with the satellite signal. To overcome this artifact, different sources close to zenith could better characterize these angular ranges.

The low far-out sidelobes in these beam patterns is attributed to the "clear aperture" of the ATA offset Gregorian telescope design. This is important because the radio sky is filled with undesirable sources of radio frequency interference, such as satellites, that are 60 – 90 dB brighter than astronomical sources. Even at the ATA, it is impossible to observe in the middle of satellite transmission bands. Experience shows that satellite transmissions tend to fall rapidly (e.g. by 60 dB over a fractional bandwidth of 1%) as the frequency tuning is changed from the main transmission band. This is helpful for astronomical observations near satellite transmission bands.

High dynamic range beam patterns can be obtained only on very strong sources such as satellites (in this case, XM radio and SatMex 5). In fact, XM radio (2335 MHz) was so strong that we had to introduce a 10 dB attenuator between the LNA and post-amplification module to perform the measurements to prevent signal compression when the telescope was pointed at the satellite.

In the vertical polarization (upper plot in Figure 9), a peculiar peak is observed at approx. 10° from boresight. This peak is associated with the "dog's eyes" pointed out in Figure 4. The gray and dashed curves in the same graph show that this feature is not a strong contributor on the horizontal polarization using the ATA feed or using an alternative narrow band corrugated horn feed (RPA Feed). At higher frequency a strong peak in the vicinity of the dog's eyes is also not apparent in the horizontal polarization (Figure 9, lower).

Further studies are needed to fully characterize the dog's eyes feature. A 30° wide beam pattern takes 6 hours at 2335 MHz and 3 days at 7575 MHz using a 1 second integration time. Also, we don't know if signals received through the dog's eyes will add coherently or incoherently in array imaging -- do all antennas have the same phase in these directions. Such studies are considered important and should be pursued in future work.

### F.    Pointing and Squint

Each ATA antenna is fitted with an 8-parameter pointing model using the TPOINT package [26]. Mis-pointings and surface imperfection depend on the weather (cloudy/clear) and there is no simple way to precalculate the effects and take them out in calibration. Later in the paper we characterize solar induced dish deformation under worst case conditions (Figure 18 and Figure 19). Since these effects cannot be predicted, we may need to make multiple pointing corrections each day to minimize solar effects on data quality. This problem has not yet been tackled by the telescope observing team, partly because solar mispointing and deformation are negligibly small for most of the observations currently pursued at the ATA.

During the daytime, direct solar illumination of the antenna pedestals has been observed to cause the antennas to tilt away from the sun (anti-sunflower effect, Rick Forster, private communication). Also, deformations of the dish (see section II.J) can influence the pointing of antennas, causing them to deviate from nighttime pointing models (pointing models are typically generated at night to avoid such time-variable effects). To demonstrate the quality of nighttime pointing,

Figure 10 shows measurements on 32 antennas for a single polarization (usually the X-pol unless that is not available). Because of polarization squint (Table 1) it is not physically possible to have a pointing model that is optimal for both polarizations. From Figure 10 we calculate the RMS pointing accuracy at nighttime to be 1.5', which is 1/14 the primary beam diameter at 10 GHz, our highest frequency.

To give an idea of the difference between nighttime and daytime pointing, Figure 11 shows an experiment performed with a small group of well-calibrated antennas at night and during the day. All the pointing results are derived from the average of many measured pointing offsets over many sources in the sky. We observe ~3' offsets from our pointing model during the day, presumably arising from sunflower motion and dish deformations.

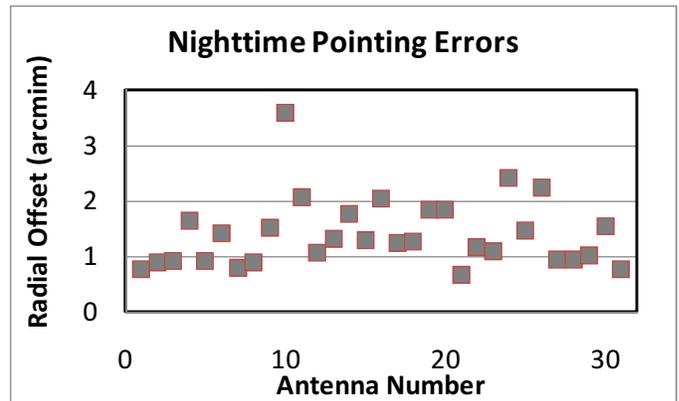

Figure 10: Nighttime measurements of the mispointing as measured on a GPS satellite for 32 antennas on x polarization. From these data an RMS nighttime value of 1.5' is determined. It is physically impossible to simultaneously apply optimal pointing models for both x- and y-polarizations, so we have arbitrarily chosen to optimize pointing for horizontally polarized radiation.

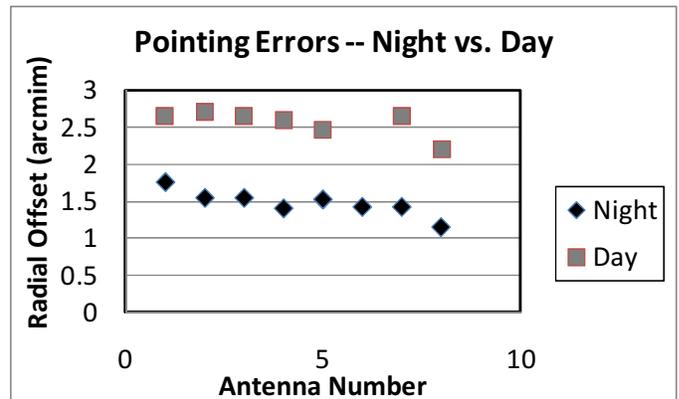

Figure 11: Comparison of daytime and nighttime pointing for selected antennas (x-polarization). From this data we estimate the daytime pointing reliability to be about 3'. Two factors reduce pointing accuracy during the day: antenna pedestal anti-sunflower effect and solar induced dish deformations (described in text).

For good mosaicking and especially for polarization studies of astronomical sources, it is important to make empirical measurements of the squint, which is defined as the angle between the pointing directions for the X-and Y-polarizations on each antenna (X = horizontal, Y = vertical polarization). A table summarizing the current squint on most antennas is displayed in Table 2.

The optics and beam patterns are mostly repeatable, so we expect similar repeatability for squint if it arises in the beam optics. On the other hand, the receiver uses two balanced amplifiers for each polarization, so if there are slight differences in impedance between the two receiver inputs for a single polarization, this will introduce squint that is not reproducible from antenna to antenna. For example, we have observed cases where one of the feed pennants broke free from the amplifier, giving an open circuit on one side of one polarization. In such cases we always observe a pointing offset of about 1°. This extreme case gives a scale against which to compare observed squint values.

|  | # dual-pol Ants | Squint < 5' | Squint 5-10' | Squint >10' |
|---|---|---|---|---|
| Before retrofits | 18 | 22% | 34% | 44% |
| After 75% of receivers were retrofitted | 32 | 37% | 42% | 21% |

Table 2: Squint (percentage of dual-pol ants & average squint in arcmin before/after retrofits).

Until a few months ago (Table 2, before retrofits), the median squint angle on our feeds was random in both magnitude and direction, with a median value of ≤10'. Presently the ATA receivers are undergoing upgrades of the pennant/amplifier connections which improve our squint and feed reliability. With the 24 dual polarization retrofitted feeds available on one day, we show the measured squints in Table 2 and Figure 12.

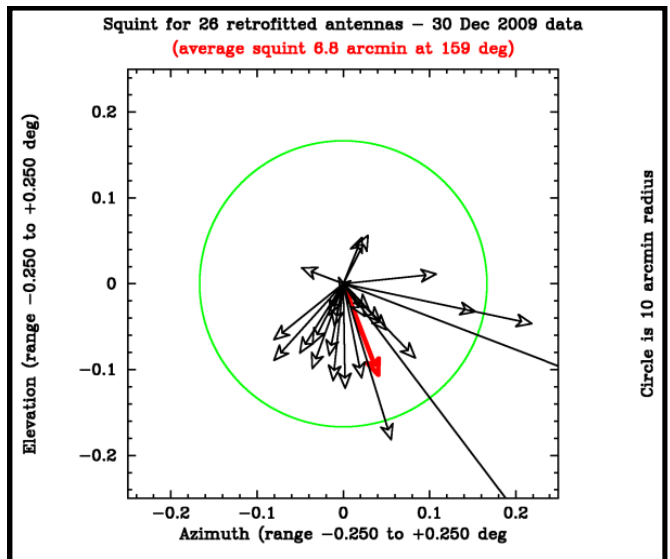

Figure 12: Vector plot of squint values from 24 retrofitted antennas.

In Figure 12 the median squint is ~4' (at 1.5 GHz, measured in single-dish mode on GPS satellites), with a few outliers below and above this value. Based on historical experience, the two largest outliers are still most likely due to a poor connection pennant and amplifier (not all feeds have undergone retrofit). There is a noticeable "average" squint over the entire data set of about 7' in the direction of lower elevation. Specifically, the Y-polarization tends to point lower than the X-polarization by this amount. Since this offset has a repeatable component despite feed differences, it is probably due to optical misalignment of the feed (e.g. feed droop). One referee correctly suggested that if e.g. the two x-pol feed pennants at top and bottom were misaligned, this will lead to a pointing offset. Microscopic inspections of a random selection of feeds shows that the top/bottom or side/side pennant offsets are no greater than 0.1 mm. This could lead to the magnitude of pointing offsets observed here at the highest frequencies, but not at the frequency where the squint observations were performed (1575 MHz).

We conclude that the contributions from optics and feed are on the same order. We note that optical squint is expected to be constant with frequency while receiver squint is more likely to be frequency dependent. We have not yet performed measurements as a function of frequency so this statement is speculative. As for squint, the present report is a work in progress.

The squint values reported here can be compared with those of other telescopes by comparing the ratio of squint to beam size. For example, at ATA the squint is 4% of beam size at 1420 MHz, at Arecibo (Arecibo, Puerto Rico) it is about 1% [27], and at the VLA it is about 3% [28] at the same frequency. At the start of the project (Table 1) the ATA had no squint specification. As far as we know, the SKA preliminary design specifications do not quote a squint value, yet.

### G. Focus

The ATA uses a unique, log-periodic feed (Figure 13) and wide bandwidth receiver that allows observations from 0.5-11.2 GHz. The log periodic structure of the feed is evident in the gradually reduced scale of the tines as they approach the feed tip. Because the active area of the feed moves with frequency, optimal results are obtained when this active area is moved into the focal plane of the dish (maximal in-out movement ~25 cm). The self-similarity of the feed results in self-similar behavior as a function of frequency.

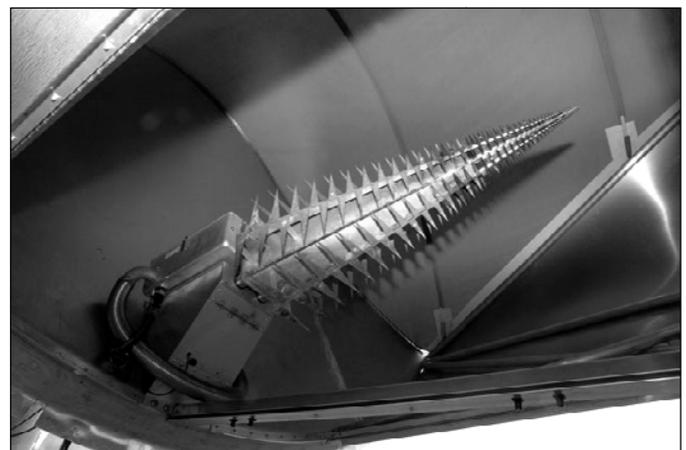

Figure 13: ATA wide-bandwidth feed mounted on one of the dishes. Photo credit, Seth Shostak.

The axial feed position has a functional relationship with the observing frequency that is optimized. Hence we define the concept of focus frequency $f_{focus}$: the frequency at which the pertinent (active) section of the feed is in the actual antenna focus plane, that is, what frequency is currently in focus.

$f_{focus}$ is distinct from the observing frequency $f_{Obs}$ since the latter depends on the tuning of the downconverter electronics and not on the feed position. To optimize point source sensitivity at the center of the FOV for a single value of $f_{Obs}$, one sets $f_{focus} = f_{Obs}$. If one uses the telescope to observe at multiple frequencies, a compromised focus frequency is chosen.

We wish to empirically verify our "focus equation" which identifies the optimal feed position for a given $f_{Obs}$. We do this by plotting the measured power on a point source over a narrow bandwidth, as a function of focus position at different observing frequencies (1575, 3877, 7575 MHz). These data are then plotted on a frequency-normalized scale $f_{focus,Norm} = f_{focus} / f_{obs}$ (Figure 14), where they should all have the same shape based on the self-similar physical shape of the feed. Figure 14 clearly shows that when using a normalized frequency scale, all curves lie on top of one another, highlighting the self-similar frequency dependence of focus for the ATA feed.

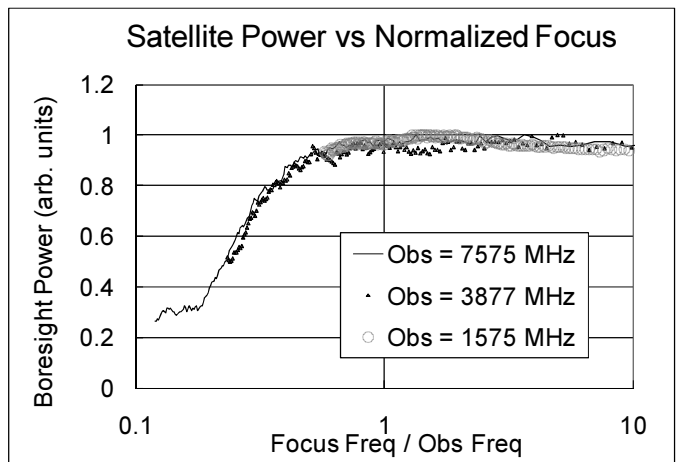

Figure 14: Boresight power as a function of feed position in units of focus frequency for 3 observations on (circles) GPS 1575 MHz, (triangles) Satmex6 3877 MHz, and (dashed line) DSCS-III 7575 MHz).

Figure 14 also brings out an important point regarding the ATA feed. We can allow the telescopes to be set *out of focus* and still achieve good performance on boresight. This is an important quality for a telescope like ATA where we intend to observe at multiple frequencies simultaneously. As long as $f_{Focus}$ is within ½ to 10 times $f_{Obs}$, de-focus has little effect on sensitivity. Since the ATA has 4 independent, simultaneous tunings, this quality is necessary for making simultaneous observations in several bands.

While Figure 14 characterizes the boresight focus properties, imaging observations require that the beam pattern has weak focus dependence (or at least that the focus dependence can be characterized). A set of beam patterns for the same satellite and observing frequency 3877 MHz is shown in Figure 15, taken as a function of focus frequency. Visual inspection of this figure shows that reasonably sharply peaked beam patterns are obtained for any focus > 3030 MHz, in alignment with the boresight power curve (Figure 14, triangles). The beam shape changes from elongated vertically at lower frequencies to elongated horizontally at higher frequencies, which indicates a small amount of astigmatism in the optics / feed. See the discussion section for more on this topic.

Performing high fidelity mosaicked imaging simultaneously at multiple frequencies may require empirical measurements of the beam pattern versus $f_{focus} / f_{obs}$. Fortunately, this focus-dependent characterization need be measured at only one observation frequency, thanks to the self-similar behavior of the feed.

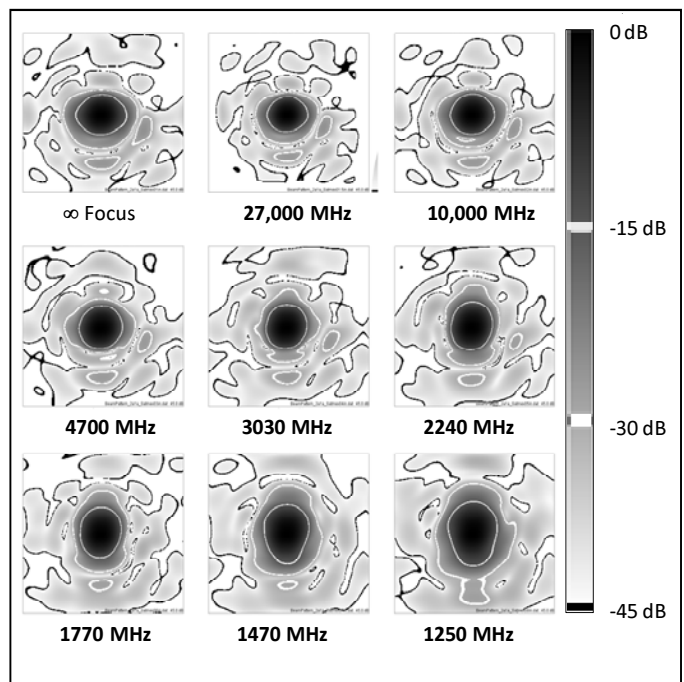

Figure 15: Beam patterns as a function of focus frequency acquired over ±4° from boresight with observation frequency 3877 MHz on the Satmex-6 geosynchronous satellite. The beam patterns are displayed on a logarithmic scale to bring out the very weak sidelobe features with contour lines at -15 dB, -30 dB and -45 dB. The correct focus frequency is visually verified as being between 4700 MHz and 3030 MHz images.

### H.    Dish to Dish Beam Pattern Repeatability

As mentioned above, the nighttime beam patterns are reproducible to a level of 1 part in $10^5$ for a single dish under similar conditions separated by a week in time [22]. Since all dishes are stamped from the same molds, we are also interested in the dish to dish repeatability. After the focus characterization above, we made a daytime run on 12 dishes (randomly chosen, available on that day) using the Satmex-6 satellite at 3877 MHz and optimal focus. The resultant beam patterns are displayed on a log scale in Figure 16. The gray scale varies over 30 dB in power, and contours are shown at -10 dB, -20 dB and -30 dB.

Of the twelve patterns, nine show striking similarities of the

-30 dB contours and a numerical computation of the weighted RMS difference across all patterns is less than 1% of the boresight beam maximum. If we number the patterns from left to right and top to bottom, outlier patterns are observed at positions 4, 9, and 12. Figure 16 lends some confidence that most antenna mirrors and their alignment are reproducible across the array. The tolerances for mirror alignment are very tight (< 1mm), so such repeatability is expected. All of the observed patterns show a high degree of similarity down to the -10 dB level.

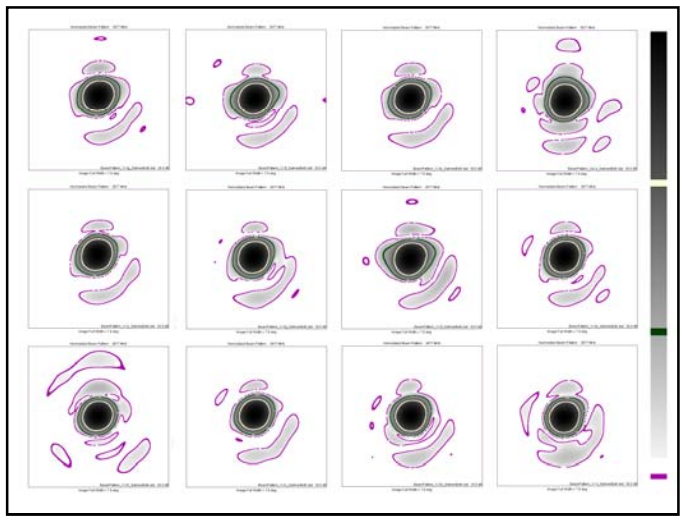

Figure 16: Grayscale / contour plots of antenna beam patterns taken over ±7.6° at 3877 MHz on 12 different antennas (X Pol only). These antennas were chosen simply based on the antennas available to us on that day. The reproducibility from one dish to the next is remarkable, as the gray color scale varies over 30 dB on a log scale. The contours indicate approximately -10 dB, -20 dB and -30 dB from boresight power.

Below -20 dB the patterns look different from one to the next, especially beyond the first two sidelobes [22]. The far-out sidelobe variation may be caused by any number of factors: real differences between the dishes including those due to non-uniform heating (see section II.J), differences in receivers, and even from scattering from adjacent dishes and infrastructure in the field (an example of this is shown in [22]). We conclude that to get the highest dynamic range, it will be necessary to use self-calibration or similar techniques to remove the contributions of strong interferers in the sidelobes of the primary dishes, at the very least. Regular self-calibration of the effective primary beam pattern may be required to get image dynamic range (not to be confused with image fidelity) on the order of $10^6$ as specified for the SKA telescope.

### I. Dish Surface Accuracy

The primary goal of holographic analysis of beam patterns is to determine the dish surface accuracy which can be characterized with a single parameter: The power-weighted RMS deviation of the dish surface from a perfectly shaped surface. Again, the coupled optical system of the ATA implies that high quality results come only when the primary dish, secondary dish, and feed are all aligned and working well.

The holography processing is described in section D above. Figure 17 gives a visual depiction of the relationship between the complex-valued primary beam pattern represented as amplitude and phase (Figure 17, left) and the image of the primary dish surface (Figure 17, right) in the plane of the fictitious symmetric dish. The dish surface shows a clean and smooth amplitude illumination across the dish. The dish illumination tapers at the edge of the dish to about 10% of the maximum in the dish illumination pattern, in good accordance with expectations based on empirical feed gain measurements[1].

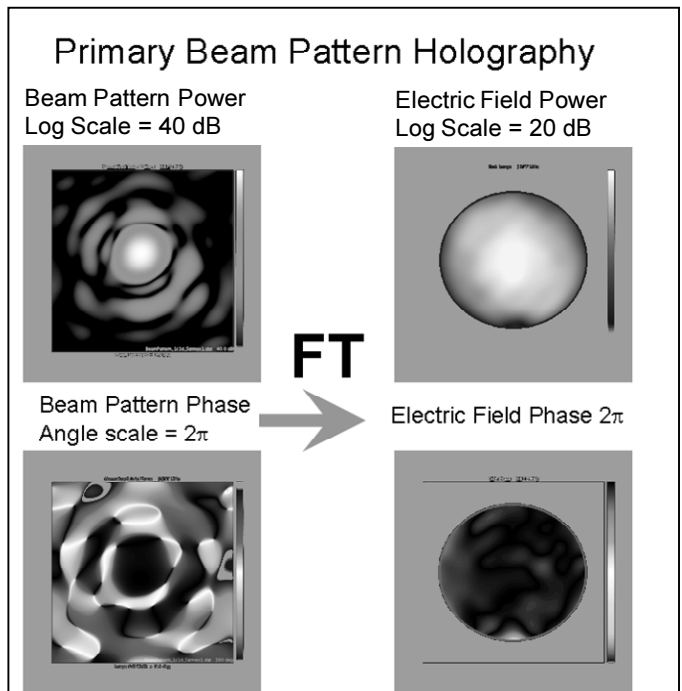

Figure 17: Schematic description of holographic process to determine the dish illumination pattern (amplitude and phase) from the measured beam pattern. The beam and dish pattern (top) are plotted on a 30 dB log scale as indicated by the color bar. The beam and dish phase patterns (bottom) are plotted on a linear scale where black represents zero phase and white represents the offset from zero phase. Note that zero phase appears at the top and bottom of the color scale because the phase wraps at 360°.

The dish pattern (bottom right) shows a maximum 10° phase non-uniformity at the edge of the dish (1/36 of the 8 cm wavelength). The inaccuracies are highest at the dish edge where they make the smallest contribution to the beam pattern. There is a dimple at the bottom of both dish amplitude and phase images which is present on every dish image. This is the shadow of the secondary dish that obscures a small section of the primary. The phase distortion in this area is high, but because of the shadowing, this part contributes negligibly to the beam pattern.

For a given point on the dish surface, the phase non-uniformity and surface roughness have a simple relationship related by the radiation wavelength: 2*(surface non-uniformity) / λ = sin(phase non-uniformity).[4] In computation of the RMS, this result is weighted by the squared magnitude of the dish illumination at the same point. From this we compute the weighted root mean square (RMS) deviation of the dish surface from that of a perfect dish. This procedure was applied to data from ~30 antennas to obtain the statistical distribution of errors from dish to dish. Measurements were made in two ways, using geostationary satellites at 3877 and 7575 MHz and using CasA as an unresolved point source at 3 frequencies where RFI is minimal: 1420 MHz, 3140 MHz, and 6280 MHz. Although not specifically planned this way, our project was allocated daylight observation time on the CasA runs and night observation time for the satellite runs.

---

[4] Caveat, the measured surface roughness contains contributions from the secondary non-uniformity, the feed, etc..

Because of the relatively low sensitivity of a single ATA dish to CasA, the CasA runs[5] had signal to noise ratios (SNR) of ~10, while the satellite SNR were $10^5$. Antenna-averaged beam patterns on CasA (1420), Satmex-6 (3877) and DSCS-III (7575) are displayed in Figure 6.

The RMS dish deformation showed excellent consistency. At night, 47 different antenna / polarizations measured at 3.8 GHz gave a median surface accuracy of 0.7 mm. Recall that before plotting, these surfaces have been corrected for mis-pointing or squint in the analysis stage. If these variations arise only in the optics and have nothing to do with the feed, then they should be frequency independent. To check this, a second series of measurements were made at 7.6 GHz, and indeed the median result was almost exactly the same. The daytime measurements had a wider spread of surface RMS values and averaging results obtained at 1.42, 3.14 and 6.2 GHz resulted in a daytime RMS value of 3 mm. Comparisons of dish deformation between day and nighttime can be found in Figure 18 and Figure 19.

The most important result is that we achieve nighttime surface median RMS values of 0.7 mm for all measured dishes. This is an important confirmation of the dish hydro-forming process and the unique support structure of the ATA antennas. The primary reflector is supported at its rim by a fan of aluminum struts to a central fixture behind the dish surface. The only other support is a flat spring supporting the center of the dish. The spring fixes the dish position both horizontally and vertically, but allows the dish to freely change in radius as it expands during heating.

### J. Direct measurements of Solar Heating Effects

When the sun is up, we expect non uniform heating of the dish surface will affect the median surface roughness, and this is demonstrated in Figure 18 and 19. While Figure 18 shows quantitative results, Figure 19 show a more qualitative comparison which highlights the areas where deformations occur. Both figures show the measured deviation from the nighttime relaxed surface for a single dish (3877 MHz on Satmex-6).

We recorded beam patterns over 50 minute durations starting on the hour. Once all the patterns were collected, we subtracted the mean of the dish patterns from 01:00–03:00 (PDT), over which time the dish appeared to be minimally changing. The weak, long wavelength striping of the images is due to a slight pointing change between 1am and 3am, and amounts to only about 0.2 mm RMS, which is not significant compared to the daytime deformations.

There is a dramatic increase in the RMS between 07:00 and 08:00, when the sun rises above the mountain ridge in the east. (The Hat Creek Radio observatory lies in a depression surrounded by ancient lava ridges on all sides.) The RMS peaks just after 13:00 when the sun transits. Another discontinuity is observed at approximately 19:00 when the sun sets below the valley rim.

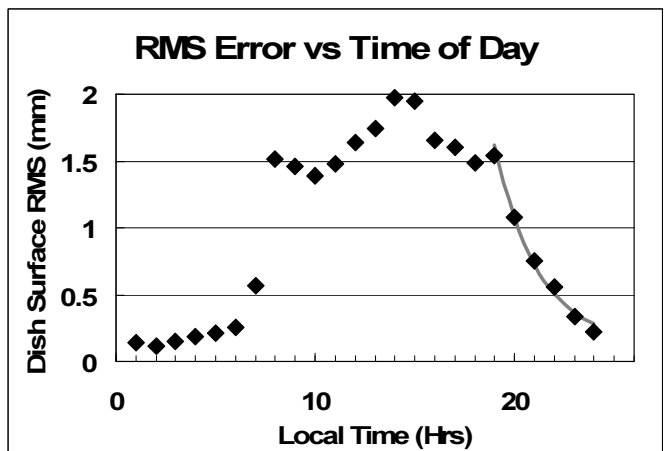

Figure 18: Relative surface error (for equivalent paraboloid) as a function of time of day. The "average dish" between 1 and 3 AM local time was subtracted from all phase images before computation of the relative differences. The dishes are more accurate at night and dish deformation RMS rises by about 2mm from night to day. On the right hand side, after sunset, the deformation is fit with an exponential relaxation curve with a half life of 1.6 hours.

To help quantify the relaxation time of the antenna after sunset, an exponential curve is fit to points between 19 and 22 hours in Figure 18, while taking into account a 0.2mm background RMS added in quadrature. The fit has an R-factor of 0.98 and indicates an RMS half life of 1.6 hours. The half life of deformations is surprisingly long, though it might be due to uneven diffuse lighting of the dish after the sun goes below the ridge.

A qualitative view of the dish deformation is shown in Figure 19, which displays grayscale images of the dish displacement from its relaxed state shape as a function of time. The sunrise between 07:00 and 08:00 is accompanied by a small hot spot on the upper left side (west) of the dish. (Compare the 08:00 surface with the image of a real ATA dish in a comparable position at sunset in Figure 1.) The west side of the primary and east side of the secondary heat because of the dish concavity. As the sun rises so does the hot spot through 13:00 hours. During this time, the sun moves behind the dish and up toward zenith. This is possible because the primary dish points "downward" relative to boresight by 29.4°, evident in Figure 1. For this example, the sunset pattern is not symmetric with sunrise because the dish was pointed about 13° east of south. At sunset, the antenna is once again illuminated from the west, now on the back side, as the sun falls behind the surrounding terrain.

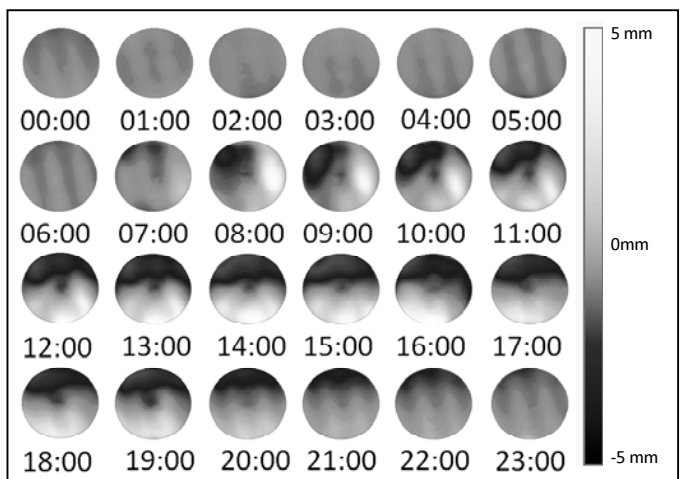

Figure 19: Grayscale display of deformation of one dish (as determined from holography) during a 24-hour period in August, 2008. Evidence of sunrise appears at ~08:00, and the deformations maximize at 13:00 hours (daylight savings time) and then decay after sunset after 19:00 hours. The dish was facing 13° east of south, hence the sunrise / sunset patterns are not symmetrical.

Another point of note is a small dimple near dish center visible on many images in Figure 19. This is at the position of the central spring support behind the dish. Finally, we have not measured or considered how dish deformations may affect squint. This may be an interesting question for a future study.

## III. DISCUSSION

### A. Ruze Law for Predicting Primary Beam Performance

The goal of this research is to validate and characterize the ATA dishes and resultant beam patterns. Important points have arisen in this study including non-circular beam shape, astigmatism, squint, and different large-angle beam pattern shapes for the two polarizations as in Figure 3. A full understanding of these effects awaits a quantitative electric field modeling of the telescope. The original EM calculations (almost a decade old) for this dish ([29]) were perhaps oversimplified and did not make a realistic calculation of contributions from line of sight between source and feed or reflections from the primary dish into the feed backlobes. For these and other reasons (perhaps dish warping during the mounting stage), the simulations did not show the asymmetries we find empirically. We are presently collaborating with the South African large MeerKAT group to generate full field ATA dish simulations which may identify the sources of the "dog's eyes," squint, and/or astigmatism.

Experimental results for the ATA show RMS errors with median 0.7 mm at night and 3 mm under worst case solar illumination (median values over a large sampling of the array). How can we extrapolate the effects of such RMS errors to give performance versus frequency? One method is to apply a Ruze model [30] for the degradation of sensitivity with increasing frequency. The Ruze model makes the assumption that the surface irregularities sample a uniform Normal distribution. This is probably not a good approximation since Figure 19 clearly shows non-uniform distribution of errors under solar illumination. Fortunately, an excellent paper by Greve [31] looks at non-Gaussian distributed deformations and divergence from Ruze law. Greve finds, for example, that as long as λ/RMS > 13 (i.e. the boresight amplitude is reduced by no more than 40%), then the particularities of the deformation distribution do not affect the boresight sensitivity – i.e. Ruze law applies. This rule is confirmed in Figure 20 where experimental points above 80% power fall close to the Ruze model. With renewed confidence in Ruze law, nighttime losses of 0.5% and <20% are expected frequencies of 4 GHz and 15 GHz.[6]

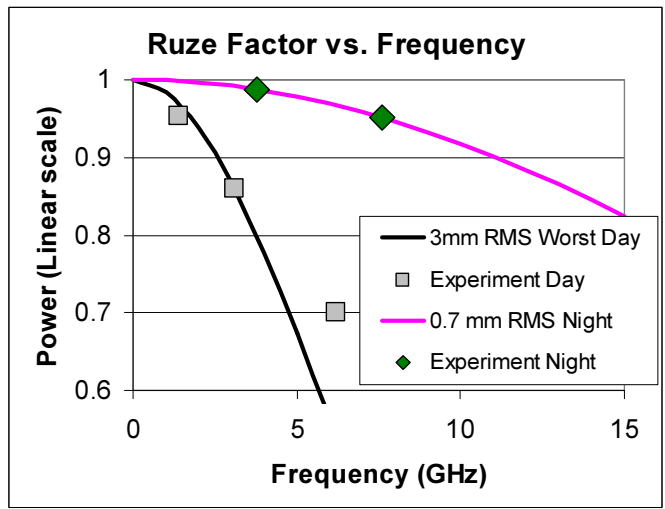

Figure 20: Calculation of Ruze factor as a function of frequency for the experimentally determined dish surface RMS (0.7 mm at night, 3 mm daytime worst case). These calculations are compared with the measured powers as derived in section 12. Night time observations may proceed up to 15 GHz with less than 20% gain loss. On sunny days, the same 20% loss limits the observations to ~5 GHz.

During the daytime, non-uniform heating of the dish can be an issue for, on average, 12 hours each day. On sunny days, solar heating can be mitigated by scheduling observations for lower frequencies (< 4 GHz). High frequency observations are best performed at night. However, solar deformations will diminish on cloudy days when the solar radiation is dispersed. The Hat Creek Radio Observatory where the ATA is located, is overcast approximately 15-20% of the days in the year. Comparisons of day time clear and cloudy beam patterns remains a task for a future study.

### B. Focus effects

In principle, the wide band single pixel feed of the ATA requires focusing, but in practice this isn't a big issue. Experiments show (Figure 14 and Figure 15) that the beam pattern is fairly stable over a decade of focus frequencies above the observing frequency. Hence, we are confident that high quality observing for point sources like SETI targets and pulsars and for areas < ½ the primary beam width can be performed at multiple simultaneous frequencies at the ATA so long as the feed is focused near the highest frequency of observation.

---

[6] Although the ATA feed is sensitive only up to about 10 GHz, there is a technology development project underway to push the frequency to 15 GHz and beyond. If this project is to succeed, we require that the dishes also operate at this frequency. Hence we include sensitivity estimates up to 15 GHz.

Another way to characterize the focus is using aperture illumination efficiencies which can be directly computed from the holography data. The "taper efficiency" measures the ratio of the integrated illumination power divided by the average illumination power of the primary dish:

$$\left( \sum_{i=1}^{N} a_i \middle/ N \right)^2 \middle/ \left( \sum_{i=1}^{N} |a_i|^2 \middle/ N \right) \quad (5)$$

Similarly, the phase efficiency isolates the effect of phase irregularities (e.g. defocus) and is defined as:

$$\left( \sum_{i=1}^{N} a_i \right)^2 \left( \sum_{i=1}^{N} |a_i| \right)^{-2} \quad (6)$$

Using data from a small range around the expected ideal focus, the taper efficiency are graphed for a 3877 MHz observation as a function of focus in Figure 21.

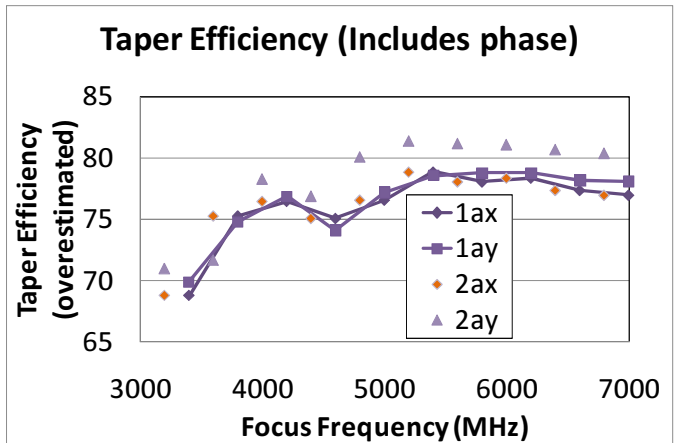

Figure 21: Taper efficiency on 2 antennas and 2 polarizations as a function of focus. Except for a small sensitivity dropout in feed response at 4.6 GHz, the taper efficiency is usually between 70-80%, at or above the observation frequency. These values reflect the same trend seen in Figure 14.

Regarding Figure 21, we make the important note that all beam patterns were measured with an angular range of about 10º, subtending only a small fraction of the entire primary beam. The reflectors have very little small-scale roughness, so this is not likely to cause a serious overestimate of the taper efficiency. In fact, a previous analytical study of the antenna properties predicted a taper efficiency of 73% [32]. For reference, the same memo predicts an aperture efficiency of 63%. The full explanation of the slightly higher measured taper efficiency values may be clarified in future full EM wave simulations. We note that computed phase efficiencies for these data are statistically indistinguishable from a hypothesis of a constant value (94%) over this focus range.

The small angular range of the beam patterns highlights information about the aperture illumination on long length scales across the aperture (as would be exhibited in a large defocus). The measured stability of the efficiency with focus change is related to the small range of feed motion (about 10 cm) for focusing as compared to the relatively large focal depth of the optics. Note that the necessary focus motion gets dramatically smaller at higher frequencies due to the log-periodic nature of the feed.

### C. Astigmatism

In beam images (Figure 15), a small amount of astigmatism is observed in beam patterns versus focus. This is quantified in Figure 22. For a sensitive ellipticity measure, we compute the ratio of horizontal/vertical beam size at the -15 dB point. Using the curve in Figure 5 we find that the beam is optimally round at ~3600 MHz, which is equal to the observation frequency within the experimental accuracy of the ellipticity measurement. Measures of the absolute beam size indicate that the astigmatism expands the -3 dB beam solid angle by <1%, which is less than the measurement accuracy of ~3%.

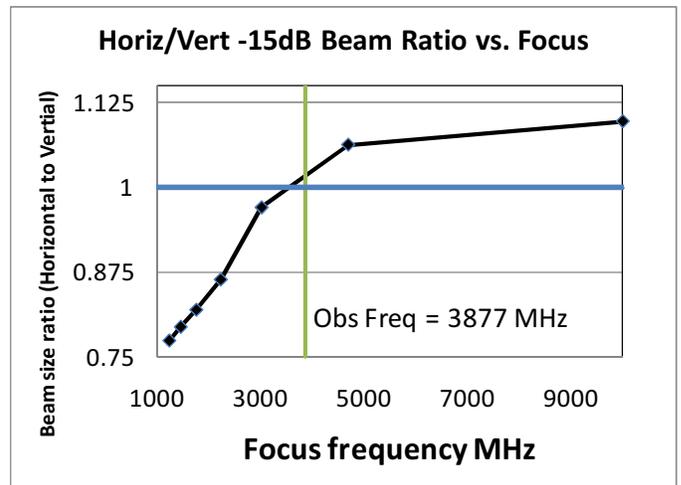

Figure 22: Measure of beam asymmetry as a function of focus position (focus frequency). Horizontal line is unity, and the experimental data cross 1 at 3600 MHz, very close to the observation frequency of 3877 MHz.

Astigmatism could be caused, for example, if the dish radius of curvature is larger on the horizontal axis than on the vertical axis, the horizontal focus will be farther from the dish and the vertical focus will be closer to dish. The feed also plays a role. The activation area of the feed for a given observation frequency is only a few centimeters (out of ~1 m), and changing the feed focus (i.e. $f_{focus}$) moves the active area to points in front or behind the focal plane. In the example above, $f_{focus} > f_{Obs}$ moves the focus closer to the dish so the vertical component is closer to being in focus. When $f_{focus} < f_{Obs}$, the beam is narrower on the horizontal axis.

Since the beam patterns and holography incorporate all effects of the feed and dish, it is difficult to disentangle the source of the astigmatism. One theory proposes that the secondary supports on the edge of the primary dish introduce a cylindrical component to the curvature on the horizontal axis. Because of the evident rectangular symmetry of the feed for a single polarization, it seems less likely that the feed produces the astigmatism (remember, astigmatism appears in a single polarization). However, if the feed droops, astigmatism may be different on the two polarizations. One way to distinguish feed contributions from dish contributions would be to rotate one feed by 90° (and perhaps 180°) and measuring beam patterns before and after. Another approach would use a full EM wave simulation of both the feed and optical elements to simulate the same factors.

Several sources of primary beam asymmetry include:

- Non-circularity of the beam pattern
- Polarization dependence (squint, feed droop?)
- Focus frequency dependence
- Solar illumination dependence
- Receiver dependence (mis-alignment or imbalanced LNA inputs)
- Dish-to-dish variations of far-out sidelobes (those below -20dB)

A complete characterization of all dishes and polarizations at all frequencies and all foci would take decades to complete (though it is possible to do that characterization for a specific frequency case). Unfortunately, solar deformations are probably not susceptible to empirical calibration since they are dependent on both weather and time of year. We conclude that in the general case, to get the highest dynamic range at ATA, *it will be necessary to use self-calibration or similar techniques* to characterize the primary beam pattern both in the field of observation and in the directions of strong interferers.

This conclusion is also relevant to the Square Kilometer Array (SKA) which may use offset secondary 15 m dishes in the frequency range 1-10 GHz (SKA hi) [33]. In many ways, the ATA is a best-case scenario for dish-to-dish reproducibility in a minimal cost design because of its molded dishes and their small size. If the ATA does not have uniform far-out sidelobe patterns from dish to dish, then it appears unlikely that a larger dish at the SKA will be much better in this regard, unless they use a different technology with much greater reproducibility from dish to dish.

### D.  Spillover Measurement

The ATA uses a unique "shroud," or shaped aluminum reflector that breaks the line of sight between the feed and the ground. Since the ground has a temperature close to 300K while the sky temperature is only 3-4 K on average, this shroud improves the system performance by removing contributions from the ground to system temperature. This point is strongly evidenced in Figure 23.

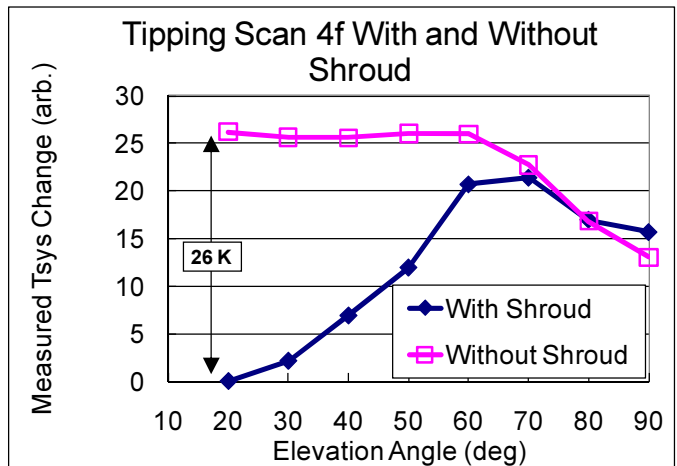

Figure 23: Tipping scan of one telescope both before and after installation of ground shield (shroud) on one antenna (4f). The shroud makes an important reduction in system temperature (compared to median receiver Tsys of 60K at 1420 MHz. Multiple scans on different antennas showed variations of ~1% between antennas and day or night.

Without the shroud, the worst case ground spillover increases the system temperature by 26 K. With the shroud at lower elevations, we suppose almost all spillover goes to the sky ($T_{sky}$ ~4 K, not including spectral line emission like HI). Near zenith, the ground spillover over the top of the dish becomes an issue independently of the shroud with a peak of ~ 20 K $T_{sys}$ increase at 70° elevation.

Another way to estimate the ground spillover at zenith is to examine the illumination power at the edge of the primary (1-3% as measured by holography). From this value and assuming a Gaussian taper beyond the dish edge, one computes a feed spillover of <11% which is consistent with computed values of primary spillover from analytic calculations [32]. Only half of the primary spillover can reach the ground even at zenith due to the shroud. If we pessimistically suppose that at zenith, all spillover radiation reaches the ground, we obtain 17K spillover noise. We do not know, but suspect a bit more ground spillover near zenith might due to line-of-sight spillover from the feed, over the top of the dish and onto the ground. All told, the spillover as a function of elevation angle (Figure 23) is consistent with measurements of the aperture illumination at the edge of the dish. Numerical simulations including realistic models of feed and optics may be the best way of clarifying the measured system temperature behavior with elevation, in future studies.

### E.  Ability to correct Primary Beam Patterns in Current Reduction Software used at ATA

At present, most ATA data are reduced using the MIRIAD astronomical processing package which supports only symmetrical, Gaussian beam shapes. The software needs to be modified to accommodate the results of empirical studies which show substantial deviations from a Gaussian beam shape. We suspect that a parameterized model of the antenna beam pattern may make a useful approximation useful in $10^4$ dynamic range mosaic observations. The construction of this parameterized model remains a task for the future.

## IV. CONCLUSION

In this paper, we document the characteristics of the ATA primary dish quality, reproducibility, and behavior under various focus and solar heating conditions. We conclude that the ATA dishes perform well enough to meet the science goals of the ATA. The offset Gregorian design with its clear aperture shows good sidelobe performance for interfering sources far from the main pointing direction (Figure 9). Results on focus properties show that our original plan of performing simultaneous high quality observations over a decade in frequency range is supported by the current system.

We have characterized the optical system performance in daylight conditions and placed upper bounds on which frequencies may be observed during the day and which frequencies are better scheduled at night. We assess the variability and predictability of ATA dish beam patterns from one dish to the next. While the primary beam is reproducible to ~1% level, time dependent deformations and far-out sidelobes are not easily predictable. This has important ramifications for data reduction both at the ATA and for the future Square Kilometer Array or any future telescope. We believe that optimal image fidelity will be achieved only when primary beam self-calibration is included in interferometric analysis. It may be possible that a different design could lead to more predictable/systematic performance, but the authors suggest that without greatly increasing telescope costs per square meter of collecting area, future telescopes may face similar issues.

In the discussion section we outline a variety of non-reproducible (from antenna to antenna) features such as far out sidelobe patterns. Beam ellipticities and polarization asymmetries, and time of day variations also affect ATA primary beam patterns. The causes of these features may be illuminated by a planned future study using full EM wave simulations of the entire optical system, including feed illumination patterns, squint / feed droop, optical asymmetries, scattering from the ground shield, etc. We emphasize that while there are still many unanswered questions, the array functions just as it is to generate high-quality science data. Finally, we present a table containing some key features of the primary dish/array performance that may be useful for future users of the array:

| Freq (MHz) | FWHM Beam Width (deg) | Receiver Noise Temp (typical) | Squint | | |
|---|---|---|---|---|---|
| 1575 | 2.22⁰ | 53 K | ~4' | | |
| 3877 | 0.9⁰ | 75 K | | | |
| 7575 | 0.46⁰ | 92 K | | | |
| 10000 | 0.35⁰ | 200 K | | | |
| | Side-lobes 3 FWHM from center | Side-lobes 20 FWHM from center | Spill-over Noise Temp El=20 deg | Spill-over Noise Temp El=70 deg | Pointing Accuracy |
| All Freqs | -35 dB | -45 dB | >~ 4K | ~20K | ~1.5' night, 3' day |

Table 3: Summary of key results which may be useful for users of the ATA.

## ACKNOWLEDGMENTS

The authors wish to especially thank all the ATA team members for unflagging encouragement and outstanding technical support throughout the 7 year work period of this report. We also gratefully acknowledge the careful reading and constructive comments of the anonymous referees, and careful proofreading by Dr. Céline Pinet.

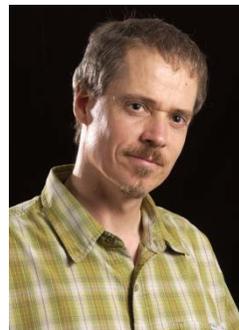

**Gerald R. Harp** (M'09) Received B.S. degrees in mathematics and physics at the University of Wisconsin-Milwaukee (UWM, Milwaukee Wisconsin) in 1986, M.S. in physics (UWM) in 1988, and Ph.D. in physics (UWM) in 1991. After a postdoctoral appointment at the IBM Almaden Research Center in San Jose, CA, Harp joined the faculty in the physics department at Ohio University in the field of solid state physics. There Harp was awarded an NSF Career grant, and promoted to associate professor with tenure. In the late 90's he joined Computer Graphics Systems Development, in Mountain View, CA, a virtual-reality startup. In 2001 he joined the SETI Institute in Mountain View, CA, where he now works as Astrophysicist. Harp is first author or co-author on more than 80 peer-reviewed papers in physics and engineering, and has authored a number of substantial memos in the Allen Telescope Array memo series (http://ral.berkeley.edu/ata/memos/). Harp has published in fields including surface science, ultra thin film superlattices, electron scattering, photoabsorption and photoemission from surfaces, electron microscopy, electron holography, optical holography and more recently radio holography. Present interests include calibration and imaging in radio astronomy, the search for extraterrestrial intelligence and the Allen Telescope Array, a radio interferometer in Northern California.

Dr. Harp is a member of the IEEE, American Physical Society, and American Astronomical Society.